# Proposed large scale monolithic fused silica mirror suspension for 3rd generation gravitational wave detectors

A. V. Cumming [1], R. Jones[1], G. D. Hammond[1], J. Hough[1], I. W. Martin[1], S. Rowan[1].

[1] SUPA[+], Institute for Gravitational Research, School of Physics and Astronomy, University of Glasgow, Glasgow, G12 8QQ, UK

+ Scottish Universities Physics Alliance

[*]Corresponding author: email alan.cumming@glasgow.ac.uk Telephone 0141 330 8237 Fax 0141 330 6833

Thermal noise from the suspension fibres used in the mirror pendulums in current gravitational wave detectors is a critical noise source. Future detectors will require improved suspension performance with the specific ability to suspend much heavier masses to reduce radiation pressure noise, whilst retaining good thermal noise performance. In this letter, we propose and experimentally demonstrate a design for a large-scale fused silica suspension, demonstrating its suitability for holding an increased mass of 160 kg. We demonstrate the concepts for improving thermal noise via longer suspension fibres supporting a higher static stress. We present a full thermal noise analysis of our prototype, meeting requirements for conceptual 3rd generation detector designs such as the high frequency interferometer of the Einstein Telescope (ET-HF), and closely approaching that required for Cosmic Explorer (CE).

*Introduction* - Advanced 2nd generation interferometric gravitational wave observatories, including Advanced LIGO (aLIGO) [1] and Advanced Virgo (AdV) [2], have regularly acquired signals from astronomical events following on from aLIGO's first binary black hole inspiral detection in 2015 [3]. To date, over 40 confirmed black hole mergers have been recorded [4], including one resulting in an intermediate mass black hole [5], and two neutron star inspirals [6,7]. Most recently two black hole/neutron star inspirals were observed [8], providing new scientific insight.

Mirrors in the interferometer's arms have final stage suspensions constructed from fused silica mirrors, each hanging from four precisely manufactured silica fibres. This is key in providing appropriate levels of noise at low frequencies (10-30 Hz), as it allows the optimisation and minimisation of thermal displacement noise [9], $x(\omega)$, given by [10,11]

$$x(\omega) = \sqrt{\frac{4k_BT}{m\omega}\left(\frac{\omega_o{}^2\phi(\omega)}{\omega_o{}^4\phi^2(\omega)+(\omega_o{}^2-\omega^2)^2}\right)}, \quad (1)$$

where $T$ is the temperature, $m$ is the pendulum mass, $\phi(\omega)$ is the mechanical loss of the pendulum mode, of resonant angular frequency $\omega_o$ , $k_B$ is Boltzmann's constant and $\omega$ is the angular frequency of interest. Choosing ultra low mechanical loss fused silica for the suspension fibres therefore permits a significant reduction in the suspension thermal noise [12-15].

Future detectors will require a radical upscaling of the suspended mass, from the current 40 kg, to 160 kg or even larger. This is driven by a need to reduce radiation pressure noise [16]. Also, the optic's larger diameter front surface allows larger beam size, reducing coating thermal noise [17]. However, this mass upscale must also not detrimentally affect suspension thermal noise performance. Therefore, suspension design must be carefully considered, and demonstrated experimentally. In this letter we propose a set of design principles for next generation silica suspensions, highlighting their possibilities and limitations in terms of noise performance and internal mode frequencies. Employing these principles, we have forged new ground by exhibiting a first successful long term experimental prototype suspension, hanging 160 kg on detector quality fibres, laser welded to silica attachment cones. Projected thermal noise performance is analysed using finite element analysis (FEA) models. When scaled to the proposed mirror masses, performance is seen to be more than sufficient to meet that required for currently proposed conceptual detector designs.

Currently, final-stage suspension designs of aLIGO and AdV have mirrors suspended from four "dumbbell" shaped silica fibres [18], with the fibre shape precisely controlled [19] to minimise thermoelastic noise [20], which results from temperature fluctuations in the fibre [21,22]. Recent studies [23] on currently installed aLIGO suspensions have demonstrated extremely low mechanical loss, actually yielding better performance than previously anticipated. There are two distinct, but potentially complementary paths that final stage suspension systems may take for future third generation telescopes. The first is to reduce thermal noise directly, via cryogenic cooling of the mirrors and suspensions. This is an extreme technical challenge requiring changing material from silica to silicon or sapphire [24]. The second path is further upgrading silica suspensions, as will be discussed in this letter. For example, the future Einstein Telescope (ET) proposes using enhanced room-temperature silica suspensions in its high frequency interferometer (ET-HF) with the instrument envisaging use of heavier silica suspensions up to 200 kg to broaden the available frequency band to higher levels >10 Hz [25,26]. Improved sensitivity at frequencies below 30 Hz comes from a cryogenic instrument with silicon suspensions and optics (ET-LF) proposed for parallel operation with ET-HF.

Interestingly, Cosmic Explorer (CE) envisages use of very large room temperature suspensions of up to 320 kg, possibly achieving similar sensitivity to ET using longer interferometer arms, without cryogenics [27,35].

Here, we propose three core concepts for improvement of final-stage fused silica suspension performance:

*Increased mass of the suspended cavity mirrors* - A strong driver for increasing the mirror mass is reduction of radiation pressure noise, which is directly proportional to the mass [16]. In addition, Equation 1 shows that the displacement thermal noise of a mirror is also proportional to $m$. Thus, by simply increasing the mirror's mass, the thermal noise performance can be improved. Typically, a mass gain of a factor of 4-8 over the current 40 kg is envisaged for future suspensions [25], For a factor 4, this reduces noise by a factor 2, as shown in the red curve in figure 1, where fibre cross section has been scaled to maintain fibre stress.

*Increased final stage length* - The second concept is to lengthen the final stage's suspension fibres. By doubling their length to 120 cm, the pendulum mode is pushed down in frequency by a factor $\sqrt{2}$. This frequency reduction effectively moves the whole curve to the left, which results in an additional thermal noise gain, yielding figure 1's green curve.

This length increase also provides the benefit of reducing the vertical bounce mode frequency by factor $\sqrt{2}$ from 8.8 Hz to 5 Hz, opening up a greater range of frequencies in the sub-10 Hz band. One disadvantage of this change is the resulting decrease in frequency of the violin modes, which encroach further into the detection band. Our third strategy change aims to help minimise this.

*Increased final stage silica fibre stress* – Due to the violin mode frequency's direct dependence on length, the fundamental mode's frequency halves to ~246 Hz when fibre length is doubled. However, fibre diameter consistency improvements [28] allow a final strategy of reduction in the diameter of the fibre's central section, increasing their suspended stress. aLIGO has a static fibre stress of 770 MPa, but fibres have been demonstrated as having ultimate tensile strengths of up to 3-4 GPa [27]. As such, increasing the stress to 1.2-1.5 GPa is a credible suspension stress for long term use, which pushes the violin modes back up in frequency, to above 300 Hz as shown figure 1's yellow curve. Whilst the original ~500 Hz frequency of the modes cannot be fully recovered, this is a necessary compromise, with the added benefit of reducing the vertical frequency by an additional factor $\sqrt{2}$.

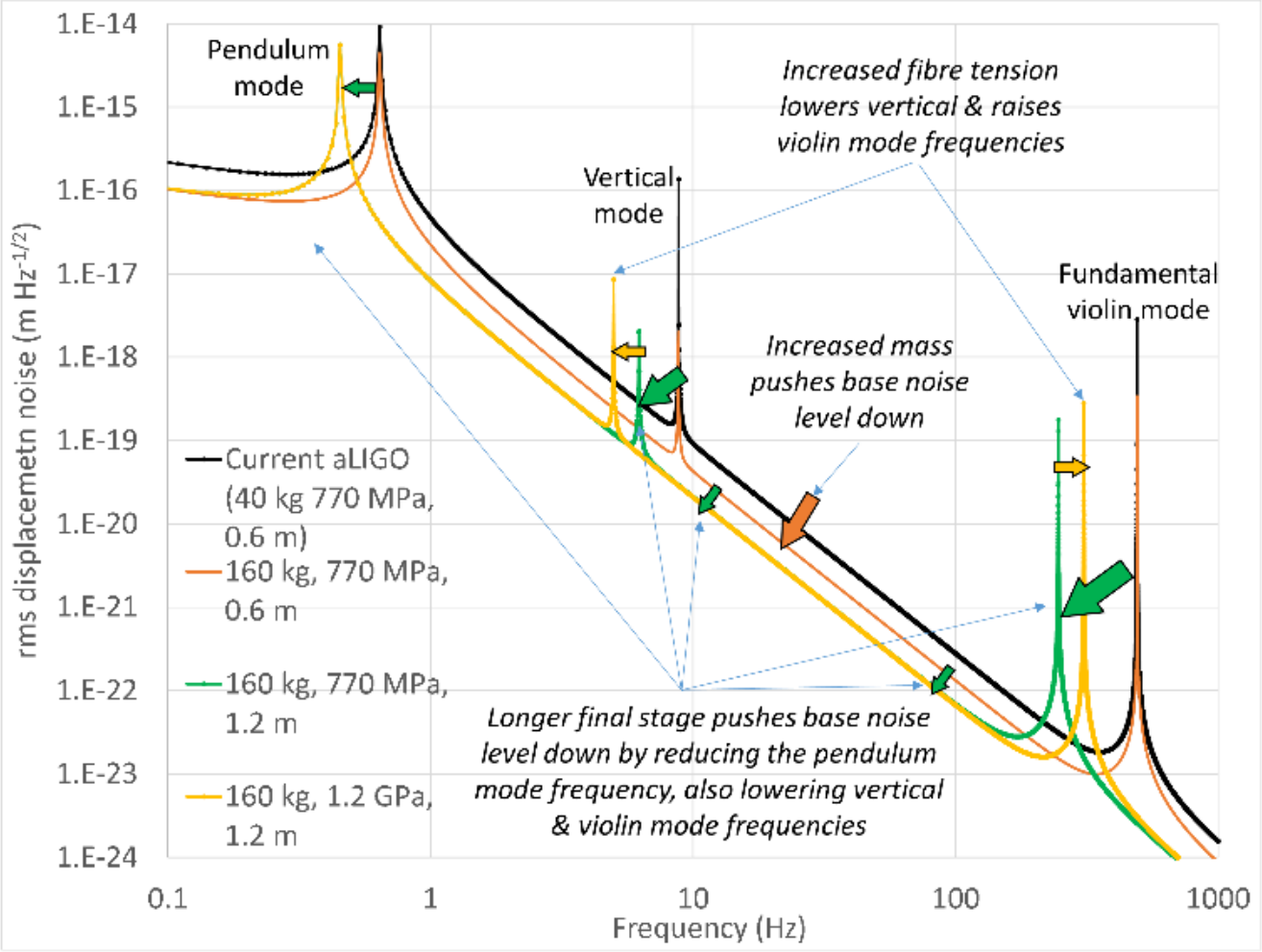

*Figure 1. Theoretical calculation of suspension thermal displacement noise of a single mirror suspension, showing the*

*sequence of steps taken to improve performance, by a. increasing the mass b. increasing the final stage length c. increasing the fibre's suspended stress*

We have successfully built the first prototype suspension employing these principles, together with a number of practical design factors which must be considered, including fibre shapes, and silica attachment piece ('ear') design, for ease of laser welding.

*Fibre shape* - The fibre design aims to follow on in concept from those used in aLIGO and AdV, via nulling thermoelastic noise that results from temperature fluctuations along the fibre, that translate into motion via the material's thermal expansion coefficient [20]. Thermoelastic loss $\phi_{thermo}$ is given by

$$\phi_{thermo}(\omega) = \frac{YT}{\rho C}\left(\alpha - \sigma_o \frac{\beta}{Y}\right)^2 \left(\frac{\omega\tau}{1+(\omega\tau)^2}\right) \quad (2)$$

with

$$\tau = \frac{1}{4.32\pi}\frac{\rho C d^2}{\kappa}, \quad (3)$$

where $Y$ is the fibre's Young's modulus, $C$ is specific heat capacity per unit mass, $\kappa$ is thermal conductivity, $\rho$ is density, $\alpha$ is the linear thermal expansion coefficient, $\sigma_0$ is the fibre's loaded static stress, $\beta = \frac{1}{Y}\frac{dY}{dT}$ is the thermal elastic coefficient, and $d$ is the fibre diameter. Nulling occurs when $\alpha - \sigma_O \frac{\beta}{Y} = 0$, giving diameter

$$d_{null} = 2r_{null} = 2\sqrt{\frac{F\beta}{\pi\alpha Y}}, \quad (4)$$

where $F$ is the vertical tension on a single fibre occurring due to the suspended mass. This yields $d_{null}$ = 1644 μm, for 160 kg mass held on four fibres. A length of 30 mm at the fibre ends was chosen to be this diameter for the nulling region, to ensure as much elastic bending energy is stored in a nulled area, with minimal contained in the starting ("stock") material at the fibre's ends used for welding to the ears [30]. Fibres for aLIGO were pulled from 3 mm diameter fused silica rod, leaving ~10 mm of this material at the fibre ends. This provided enough material to hold the fibre securely while welding occurred, while minimising the amount of elastic bending energy occurring in this region. It is important to keep the proportion of elastic energy in the weld region to an acceptable level (typically <10% in aLIGO) due to it having higher loss than the surrounding material [23].

In order to understand which areas of the fibre contribute to the dissipation, we need to consider where the elastic bending energy resides in the fibre [32]. If the mass is increased on a 3 mm stock fibre, significantly more elastic energy is pushed into the stock region of the fibre and out of the thermoelastic nulling region – as is shown in figure 2. The solution to maintain a low elastic energy content in the stock material is to increase its diameter from 3 mm to 5 mm. Figure 2 shows an ANSYS FEA case study of the energy distribution at the fibre end for 3 different cases:

*i.* an aLIGO 3 mm end diameter fibre, holding 10 kg (for total mass 40 kg),
*ii.* the same fibre holding 40 kg (for total mass 160 kg),
*iii.* a fibre with increased end diameter of 5 mm holding 40 kg.

Increasing the fibre end diameter to 5 mm gives a smaller level of elastic energy in the stock material than for an equivalent 3 mm fibre. It also results in a lower proportion (16%) of energy in the stock compared to the original aLIGO situation (25%), this being an additional improvement. The remaining 84% of the bending energy is contained in the 30 mm nulling region as shown in figure 2.

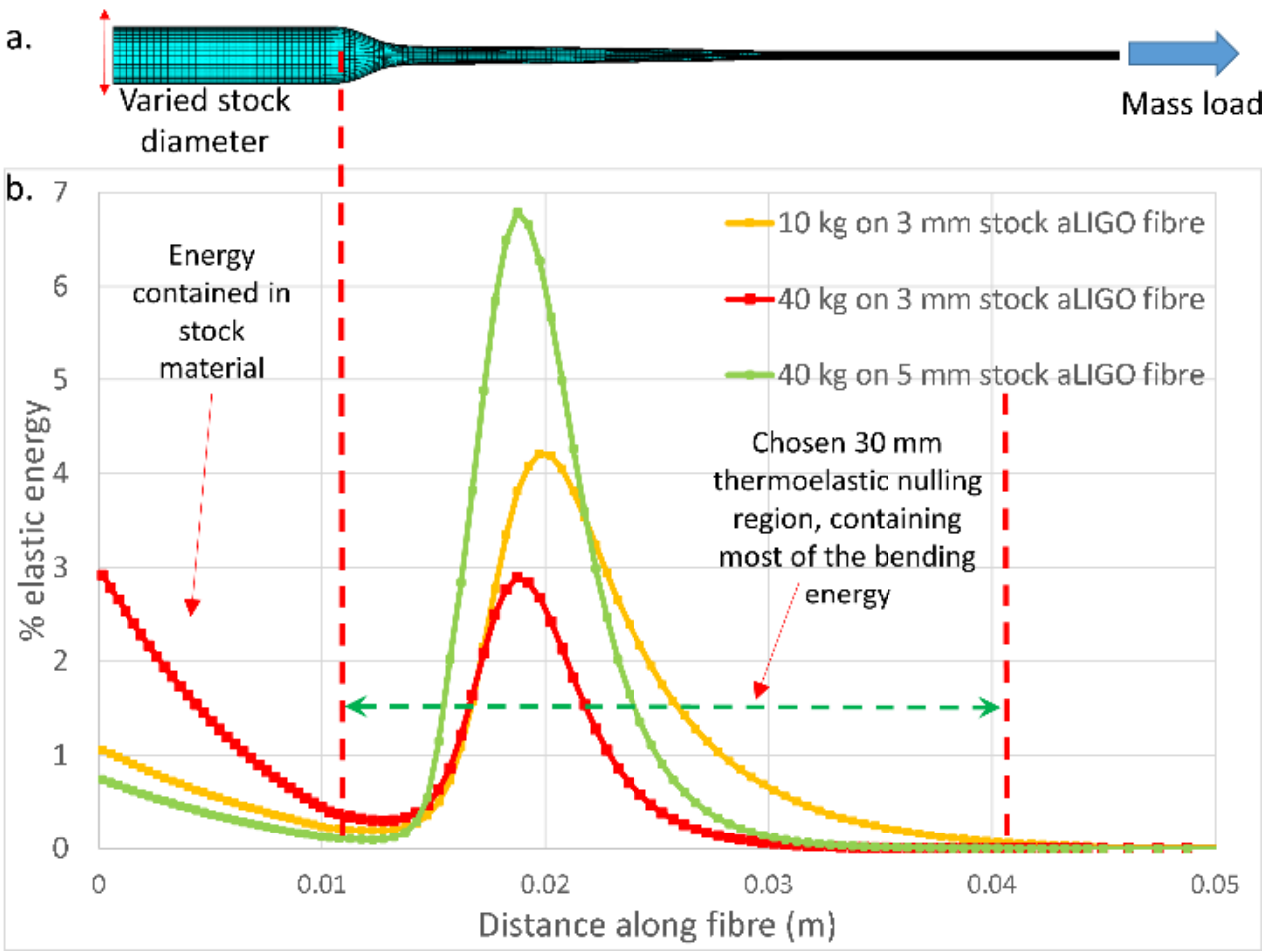

*Figure 2 a. FEA model of an aLIGO fibre, with its shorter ~20 mm nulling region b. elastic energy distribution at equivalent position along the fibre, showing results from different suspended masses and different stock material diameters, together with the proposed thermoelastic nulling region.*

The central fibre section in aLIGO is thinned to achieve optimal violin and vertical mode frequencies. For large scale suspensions, this technique will also be employed. For a 160 kg suspension, to achieve a working stress of 1.2 GPa, a fibre of diameter 646 μm is required.

Fibres were pulled from 5 mm diameter Suprasil-3 stock material which was laser polished prior to pulling to maximise strength [29]. All fibres were pulled on a lengthened version of the laser pulling machine detailed in [19], and were carefully profiled after pulling and welding on a similarly uprated dimensional characterisation machine [31] as shown in figure 3a,b, with fibres dimensions lying well within 10% of the required dimensions. The maximum fibre stress at any individual point was 1.29 GPa, and the average stress in the fibre's central section was 1.12 GPa. Their average diameter was 670 µm, approximately 4% above the nominal 646 µm diameter. Future loading tests may hang up to 200 kg, with the absolute maximum stress in this case being envisaged to be a reasonable 1.60 GPa.

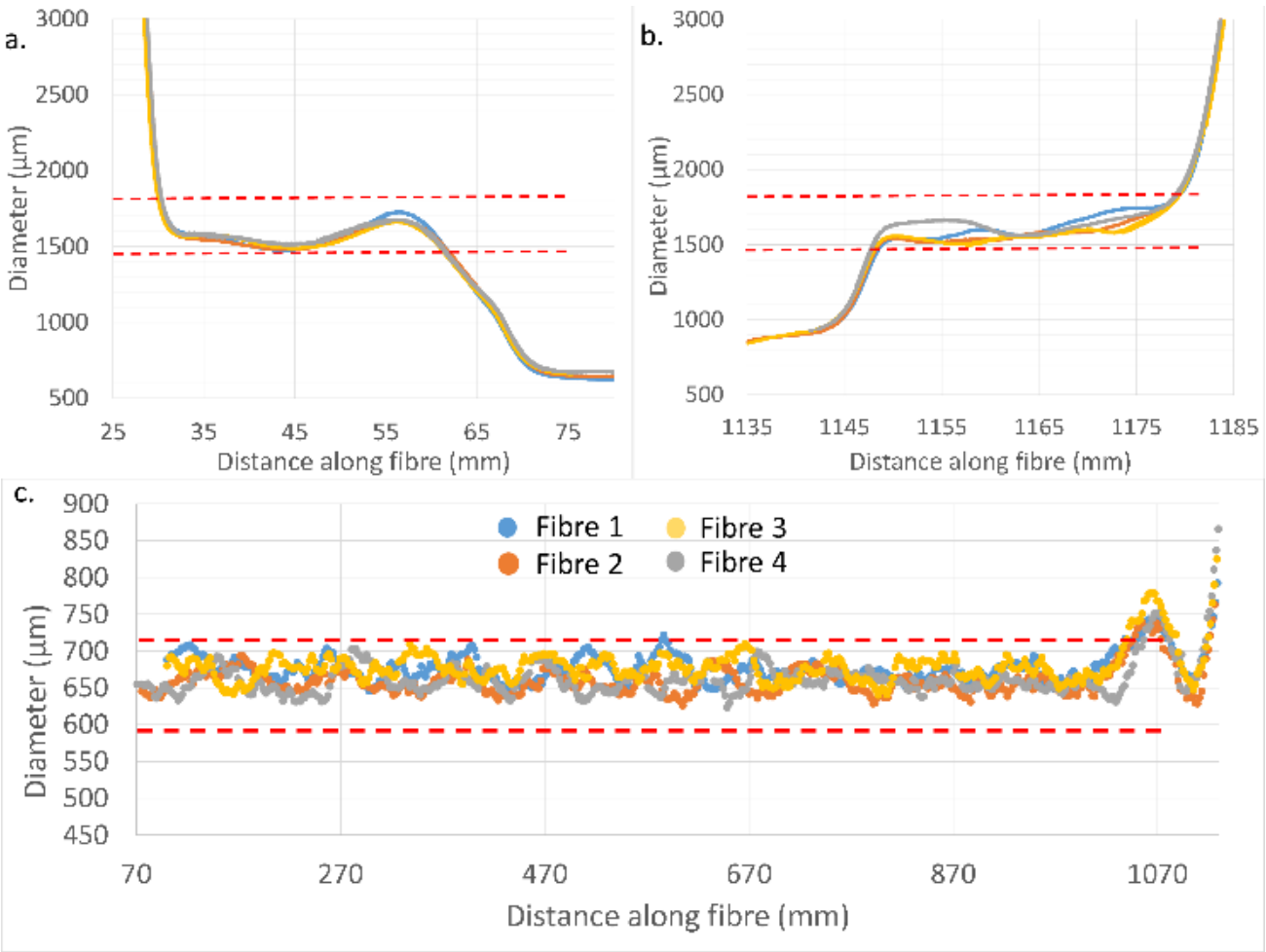

*Figure 3a. Diameter profile of bottom thermoelastic loss nulling region of fibres used in the prototype suspension b. top thermoelastic loss nulling region profile, c. central region profile. Red dashed lines show a guide to the eye of ±10% from the nominal required dimensions.*

*Laser welding* - Attachment of the fibres uses the same principle of 10.6 µm wavelength $CO_2$ laser welding that was successfully used in aLIGO. However, for thicker stock material, testing has shown that the nominal 120 W laser power available was insufficient. Therefore, a new laser of increased power of up to 450 W was used. Welding was undertaken using a conical mirror arrangement of the fibre pulling machine [19], allowing uniform heating from all sides of the fibre and ear tip ('horn'). Additional visibility was supplied using multiple 14 bit cameras [28]. This improved procedure has resulted in extremely good quality consistent welds, achievable more quickly. Research is ongoing towards a miniature portable version of this optical system for use at detector sites, to allow more amenable welding of suspensions on site at the ends of the interferometer arms.

The horn of the projected ear has been simplified from the rectangular horn with curved edges used in aLIGO, to a circular cone tip with 45 degree angle. This eases welding further by keeping both surface geometries circular at the interface point. The sharpness of the cone angle helps keep elastic energy out of the ear and deeper weld region, maintaining as much of it in the fibre as possible. The horns were clamped to bespoke metal clamps for this test, however it is ultimately envisaged that that a conical horn would be incorporated into an upscaled aLIGO style ear [30], research on which is ongoing.

The completed welded 4 fibre prototype suspension is shown in figure 4.

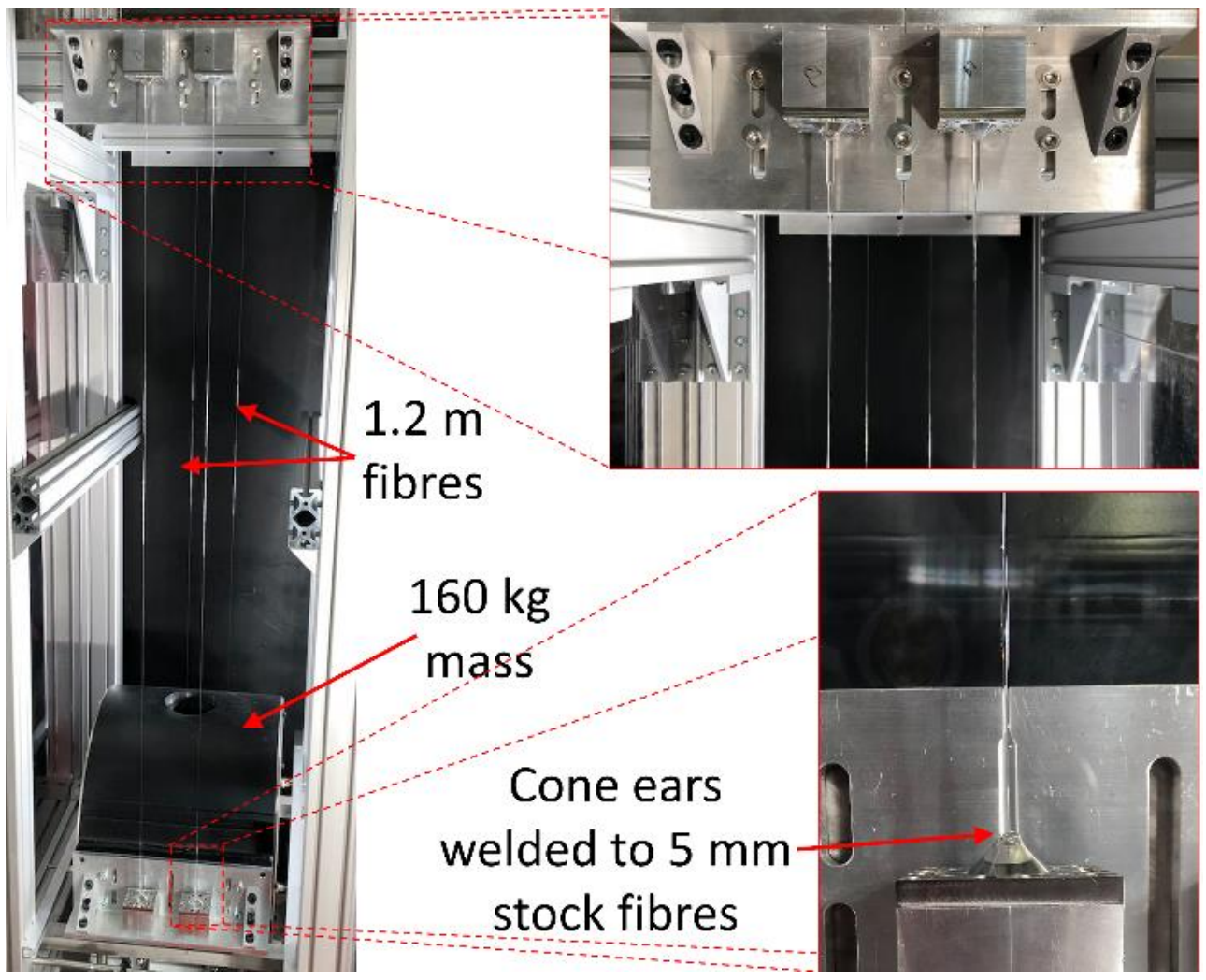

*Figure 4 Completed monolithic conceptual prototype using 160 kg mass and four 1.2 m long fibres supporting ~1.2 GPa stress. insets show 5 mm stock at ends of fibres welded to cone attachment ears and clamped to mass sides.*

Fibre profiles and welding were first tested on two single fibre 40 kg suspensions, prior to the 4 fibre full hang, with single fibre tests successfully lasting multiple months prior to the full suspension build. The four fibre prototype suspension has been hanging in standard laboratory setting in air for more than 3 years since 2018, accumulating just over 100,000 fibre hanging hours to date.

*Projected performance* – Figure 3's profile data was directly imported into ANSYS to construct an accurate FEA model of the suspension using techniques described in [30,32], here also including accurate weld geometries. The FEA model was used to evaluate the elastic energy distribution for each of the suspension's resonant modes to weight the mechanical loss contribution of each loss mechanism as previously outlined in [30].

The first mechanism is residual thermoelastic loss in sections of the fibre outwith the nulling region described by equation 2. Secondly, surface loss in the $i^{th}$ element of the fibre model is given by [21]

$$\phi_{\text{surface}} = \frac{8h\phi_s}{d_i} , \qquad (5)$$

where $h\phi_s$ is the product of the mechanical loss of the material surface, $\phi_s$, and the depth, $h$, over which surface loss mechanisms are believed to occur, and $d_i$ is the average diameter of the $i^{\text{th}}$ element along the fibre length. Using the analyses recently undertaken by the authors', $h\phi_s$ was taken to be $2.5 \times 10^{-12}$ m [23].

The third mechanism is weld loss, the excess loss associated with the welded attachment regions at the fibre ends. The value taken was $1.4\times 10^{-7}$ , again from the authors' recent analyses [23], with the observed weld region being 2 mm in length at the fibre ends [30], typically containing <5% of the elastic energy in this suspension.

Finally, bulk loss is the internal friction of the fibre material given by

$$\phi_{\text{bulk}}(\omega) = C_2 \left(\frac{\omega}{2\pi}\right)^{0.77} , \qquad (6)$$

where $C_2$ is an empirically evaluated constant [13]. For the fibre material, Suprasil 3, $C_2$=1.18±0.04 x $10^{-11}$.

The total mechanical loss is the sum of these effects, was evaluated as previously [23], using

$$\phi_{\text{resonance } j}(\omega) = \frac{E_{\text{total } j}}{E_{\text{gravity}}} \left( \sum_{i=1}^{n} \frac{E_i}{E_{\text{total } j}} \begin{pmatrix} \phi_{\text{thermo}}(\omega) \\ +\phi_{\text{surface}} \\ +\phi_{\text{bulk}}(\omega) \end{pmatrix} + \frac{E_{\text{welds}}}{E_{\text{total } j}} \phi_{\text{welds}} \right) \qquad (7)$$

where, $E_i$ is the energy stored in the $i^{\text{th}}$ FEA model element, $E_{\text{welds}}$ is the energy in the weld region, and $E_{total\, j}$ is $j^{th}$ mode's total elastic energy. $\frac{E_{\text{total } j}}{E_{\text{gravity}}}$ is the $j^{th}$ mode's 'dissipation dilution', a reduction factor occurring because a significant proportion of the pendulum's potential energy is stored in the lossless gravitational field, and is evaluated from FEA [32].

The thermal horizontal displacement noise was then evaluated from equation 1. Vertical noise was calculated via equation 17 in [30], and the violin mode thermal noise [33] was determined via equation 18 in [30]. The resulting thermal noise displacement curve is shown in figure 5.

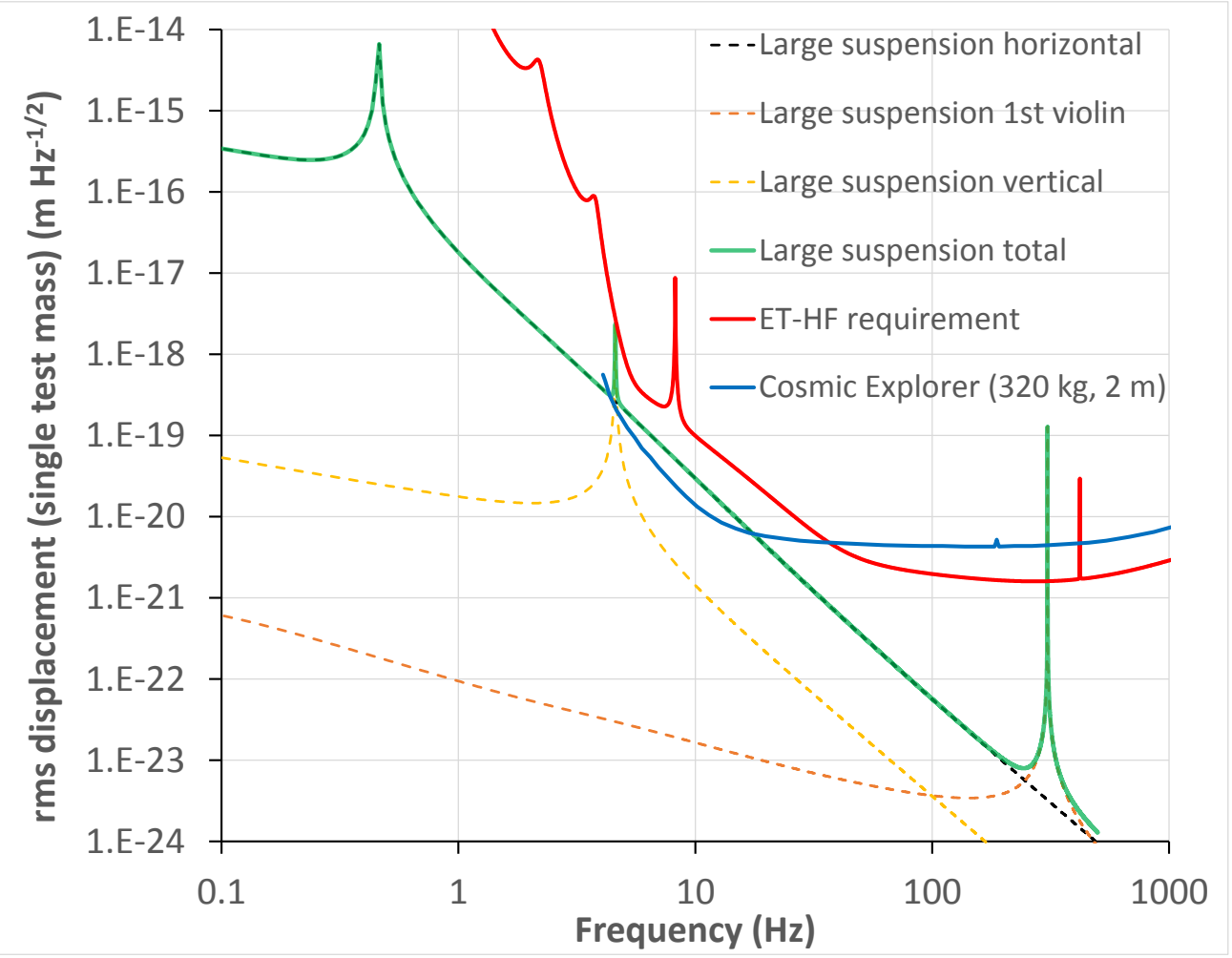

*Figure 5. Displacement thermal noise spectrum of large suspension prototype, compared to the ET-HF design requirement [34], and the CE room temperature design. CE has heavier masses of 320 kg and 2 m long suspension fibres here [35].*

ET-HF will eventually operate in combination with ET-LF, with the latter dominant below 30 Hz. However, with the technological challenge of ET-LF's cryogenics [34], it is plausible that there may be operational periods, particularly in ET's early days, when ET-HF could be running solo. ET-HF will be dominated strongly by suspension thermal noise between 7 and 30 Hz, this being some 6 times higher than the next contributing noise source, radiation pressure noise. At 10 Hz, the design requires a total noise of 9.90 x $10^{-20}$ m$\text{Hz}^{-1/2}$; this being primarily suspension thermal noise dominated, and our proposed suspension provides a considerably lower value of 2.85 x $10^{-20}$ m$\text{Hz}^{-1/2}$. Thus, deployment of the proposed suspension would yield an immediate significant gain of a factor 3.5x in detector sensitivity. This room temperature gain with proven technology could enhance any early roll out of $3^{\text{rd}}$ generation room temperature instruments, increasing astronomical reach between 7 and 30 Hz. The reduced suspension thermal noise is also less than a factor 2 above radiation pressure noise, and further gains may be achievable to make suspension noise no longer dominant using a test mass of 200 kg or higher, for example.

Cosmic Explorer's room temperature curve is an example of this, with anticipated use of 320 kg silica payloads, and fibres of length 2 m [35]. By using a simple scaling of our 160 kg, 1.2 m suspension, a mass of 320 kg would give a factor $\sqrt{2}$ improvement, and a 2 m suspension a factor $\sqrt{1.7}$ improvement, meaning a cumulative factor of $\sqrt{3.7}$=1.9. Simple scaling of our prototype shows that it approaches what would be required for CE. At 10 Hz for example, CE requires 1.33 x $10^{-20}$ $mHz^{-1/2}$ , and our suspension, simply scaled, approaches this with value ~1.50 x $10^{-20}$ $mHz^{-1/2}$. Further gain with use of even thicker stock material of up to 6 or 7 mm for higher masses. Thus, with the groundwork laid with the prototype suspension demonstrated in this letter, future research will focus on practical demonstration of a 320 kg CE capable payload, and this research is ongoing.

*Conclusions* - We have proposed 3 distinct concepts for radically upscaling final stage fused silica quasi-monolithic suspensions for gravitational wave detectors, including increasing the test mass size, increasing suspension length, and increasing stress in the fibres. We have demonstrated production of suitable fibres, and welds suitable for future detector use, and have constructed the first large scale long term quadruple fibre suspension, which has been successfully hanging for over 3 years. Projected noise performance, using FEA, has shown significantly lower suspension thermal noise than previous conceptual designs such as ET-HF, providing a potential factor of 3.5 improvement between 7 and 30 Hz, and lays the groundwork for future upscaling to 320 kg for CE.

*Acknowledgements* - We are grateful for the financial support provided by Science and Technology Facilities Council (STFC) (awards ST/N005422/1, ST/V001736/1), the Scottish Funding Council (SFC), the Royal Society, the Wolfson Foundation, and the University of Glasgow in the UK. LIGO is a facility operated on behalf of the NSF by Caltech and MIT. We would like to thank our colleagues in the LSC and Virgo collaborations and within SUPA for their interest in this work. This paper has LIGO document number P2100290.

*References*

1. Aasi, J. et al., *Advanced LIGO*, Class. Quantum Grav. 32(7), 074001 (2015).
2. Acernese F., et al., *Advanced Virgo: a second-generation interferometric gravitational wave detector* . Classical and Quantum gravity, 2014. **32**(2): 024001.
3. Abbott B. et al *Observation of Gravitational Waves from a Binary Black Hole Merger* Phys. Rev. Lett., 2016, 116 061102
4. Abbott R. et al *GWTC-2: Compact Binary Coalescences Observed by LIGO and Virgo during the First Half of the Third Observing Run* Phys. Rev. X, 2021, 11, 021053
5. Abbott R. et al *GW190521: A Binary Black Hole Merger with a Total Mass of 150 $M_\odot$* Phys. Rev. Lett. 125, 101102
6. Abbott B. et al *GW170817: Observation of Gravitational Waves from a Binary Neutron Star Inspiral* Physical Review Letters 2017,119, 161101.
7. Abbott B. et al *GW190425: Observation of a compact binary coalescence with total mass ~3.4 $M_\odot$* Astrophys. J. Lett. 892, L3 (2020)
8. R. Abbott et al *Observation of Gravitational Waves from Two Neutron Star –Black Hole Coalescences* The Astrophysical Journal Letters (2021) 915 L5–Black Hole Coalescences
9. Saulson, P.R., *Thermal noise in mechanical experiments.* Physical Review D, 1990. 42(8): p. 2437-2445.
10. Callen, H.B. and Welton, T.A., *Irreversibility and Generalised Noise.* Physical Review, 1951. 83(1): p. 34-40.
11. Callen, H.B. and Greene, R.F., *On a Theorem of Irreversible Thermodynamics.* Physical Review, 1952. 86(5): p. 702-710.
12. Rowan, S. and Hough, J., *Gravitational Wave Detection by Interferometry (Ground and Space).* Living Reviews in Relativity, 2000. 3: p. 3.
13. Braginsky, V.B., et al., *On the thermal noise from the violin modes of the test mass suspension in gravitational wave antennae.* Physics Letters A, 1994. 186: p. 18-20.
14. Braginsky, V.B., et al., *Energy dissipation in the pendulum mode of the test mass suspension of a gravitational wave antenna.* Physics Letters A, 1996. 218: p. 164-166.
15. Penn, S.D., et al., *High Quality Factor Measured in Fused Silica.* Review of Scientific Instruments, 2001. 72: p. 3670-3673.
16. S Sakata et al *A study for reduction of radiation pressure noise in gravitational wave detectors* J. Phys.: Conf. Ser. 2008 122 012020
17. Gregory M Harry et al *Thermal noise in interferometric gravitational wave detectors due to dielectric optical coatings* Class. Quantum Grav. 2002 19 897.
18. Willems P, *Dumbbell-shaped fibers for gravitational wave detectors* Phys. Lett. A 2002, 300 162–8
19. Heptonstall A et al *$CO_2$ laser production of fused silica fibers for use in interferometric gravitational wavedetector mirror suspensions* Rev. Sci. Instrum. 2011, 82, 011301.
20. Cagnoli, G. and Willems, P., *Effects of nonlinear thermoelastic damping in highly stressed fibres*. Physical Review B, 2002. 65: p. 174111.
21. Gretarsson A and Harry G M 1999 *Dissipation of mechanical energy in fused silica fibers* Rev. Sci. Instrum. 70 4081–7
22. Bell, C., et al., *Experimental results for nulling the effective thermal expansion coefficient of fused silica fibres under a static stress* Class. Quantum Grav. 31 (6) 065010 (2014)
23. Cumming A. V. et al *Lowest observed surface and weld losses in fused silica fibres for*

*gravitational wave detectors* Class. Quantum Grav. 2020, 37 195019

24. Miyoki S., *Current status of KAGRA* Proc. SPIE 11445, Ground-based and Airborne Telescopes VIII, 114450Z (13 December 2020); doi:10.1117/12.2560824
25. *Science Case for the Einstein Telescope* https://arxiv.org/abs/1912.02622
26. 2011 Einstein gravitational wave Telescope Conceptual Design Study https://tds.virgo-gw.eu/?call_file=ET-0106C-10.pdf
27. B P Abbott, et al *Exploring the sensitivity of next generation gravitational wave detectors* Class. Quantum Grav. 34 (2017) 044001
28. Lee K-H et al *Improved fused silica fibres for the advanced LIGO monolithic suspensions* Class. Quantum Grav. 2019 36 185018
29. Heptonstall A. et al *Enhanced characteristics of fused silica fibers using laser polishing* Class. Quantum Grav. 2014, 31 105006.
30. Cumming A.V. et al *Design and development of the advanced LIGO monolithic fused silica suspension* 2012 Class. Quantum Grav. 29 035003.
31. Cumming A et al 2011 *Apparatus for dimensional characterization of fused silica fibers for the suspensions of advanced gravitational wave detectors* Rev. Sci. Instrum. 82 044502
32. Cumming A et al 2009 *Finite element modelling of the mechanical loss of silica suspension fibres for advanced gravitational wave detectors* Class. Quantum Grav. 26 215012
33. Brif C 1999 *Notes on anelastic effects and thermal noise in suspensions of test masses in interferometric gravitational-wave detectors* LIGO Document T990041
34. *Design Report update 2020 for Einstein Telescope* https://apps.et-gw.eu/tds/?content=3&r=17245
35. *A Horizon Study for Cosmic Explorer Science, Observatories, and Community* LIGO Public Document P2100003